\documentclass[pdflatex,sn-mathphys-num]{sn-jnl}

\usepackage{graphicx}%
\usepackage{multirow}%
\usepackage{amsmath,amssymb,amsfonts}%
\usepackage{amsthm}%
\usepackage{mathrsfs}%
\usepackage[title]{appendix}%
\usepackage{xcolor}%
\usepackage{textcomp}%
\usepackage{manyfoot}%
\usepackage{booktabs}%
\usepackage{array}
\usepackage{float}
\usepackage{placeins}
\newcolumntype{L}[1]{>{\raggedright\arraybackslash}p{#1}}
\newcolumntype{C}[1]{>{\centering\arraybackslash}p{#1}}
\usepackage{algorithm}%
\usepackage{algorithmicx}%
\usepackage{algpseudocode}%
\usepackage{listings}%

\theoremstyle{thmstyleone}%
\theoremstyle{thmstyletwo}%

\theoremstyle{thmstylethree}%

\begin{document}

\title[Article Title]{Learning Engagement Assistant (LEA): Cross-Course Scalability and Classroom Evaluation of an Agentic AI Tutoring System}



\author[1]{\fnm{Teri} \sur{Rumble}}\email{teri.rumble@gmail.com}

\author*[1]{\fnm{Javad} \sur{Zarrin}}\email{j.zarrin@abertay.ac.uk}

\author[1]{\fnm{P. George} \sur{Lovell}}\email{p.g.lovell@abertay.ac.uk}

\author[1]{\fnm{Ruth} \sur{Falconer}}\email{r.falconer@abertay.ac.uk}

\affil[1]{\orgdiv{Faculty of Design, Informatics and Business}, \orgname{Abertay University}, \orgaddress{\city{Dundee}, \country{United Kingdom}}}

\abstract{This paper is an extension of a paper presented at the ICAART 2026 conference, which introduced \textit{LEA} (Learning Engagement Assistant), an adaptive AI tutoring agent combining course-specific Retrieval-Augmented Generation (RAG) with structured Knowledge Component (KC) models across integrated Chat, Tutor, and Quiz modes. That prior work validated \textit{LEA} on a single STEM course (CMP511) exclusively through simulation, using synthetic learner agents. This paper extends that work by reporting the first classroom deployment of \textit{LEA} with real students ($n=8$, CMP511) and the first empirical test of its cross-course scalability, deploying the system across three courses spanning two academic levels and two disciplinary domains. The study reveals a divergence from simulation predictions across modes, showing that synthetic evaluation alone cannot anticipate all aspects of real deployment. A RAGAS-based cross-course scalability evaluation (660 questions) finds Answer Relevancy and Context Precision broadly stable across courses (0.88–0.94 and 0.88–0.90 respectively), while Faithfulness declines with curriculum distance from the system's original course (0.69 to 0.50), a preliminary finding that may reflect generation logic tuned to the system's original subject rather than a scalability limitation. These findings suggest that while the orchestration layer requires no modification, full course-agnosticism of all downstream components requires further investigation.}

\keywords{AI Agents, Intelligent Tutoring Systems, Personalised Learning, Knowledge-based Systems, Retrieval-Augmented Generation, Knowledge Component Modelling, Human-Subjects Evaluation, Cross-Course Scalability}


\maketitle

\section{\uppercase{Introduction}}
\label{sec:introduction}

This article is an extended version of our conference paper presented at the 18th International Conference on Agents and Artificial Intelligence (ICAART~2026)~\cite{rumble2026learning}. That paper introduced the \textit{Learning Engagement Assistant} (\textit{LEA}) as a proof of concept validated entirely through simulation on a single STEM course, and explicitly identified real-world deployment and cross-course generalisation as necessary next steps. The present article addresses both of these deferred questions and broadens its supporting review of related work accordingly.

The landscape of AI-assisted education has shifted markedly since 2025. The prior generation of Intelligent Tutoring Systems (ITS) focused on building systems capable of personalised learning by adapting to the needs of the individual learner~\cite{latif2026systematic}. In 2025, general-purpose large language models (LLMs) acquired learning-oriented interfaces, and while vendor-published evaluations are beginning to appear~\cite{google2026learnlm_sierraleone}, independently replicated, classroom-grounded evidence of their pedagogical effectiveness remains scarce. OpenAI's ChatGPT Study Mode, Google's Gemini Learn LM, and Anthropic's Claude for Education~\cite{openai2025studymode, google2025learnlm, anthropic2025education} have joined Khan Academy's Khanmigo~\cite{shetye2024evaluation} and Harvard's CS50 AI Tutor~\cite{liu2025improving} as widely deployed conversational tutoring tools. Currently, there is a widening gap between deployment speed and evidence: these systems are entering classrooms faster than rigorous evidence of their effectiveness can be established~\cite{mannekote2024large, latif2026systematic}.

The general-purpose LLM tools share a common architecture: a Socratic-style prompting layer that guides learners toward answers rather than directly supplying them, with no course-specific knowledge grounding beneath it. What none of these tools share (including the more purpose-built systems such as Khanmigo and CS50) is a unified architecture that combines a structured Knowledge Component (KC) model, persistent mastery tracking, course-specific Retrieval-Augmented Generation (RAG) grounding, and real-time estimation of Cognitive Load (CL), Zone of Proximal Development (ZPD), and motivational state within a single system (Table~\ref{tab:its_comparison}). Mannekote et al.\ argue that without structured learner modelling (tracking knowledge state, motivation, and affect across interactions), LLM-based tutors cannot achieve the pedagogical depth required for genuine whole-learner support~\cite{mannekote2024large}. General-purpose LLM tool course awareness is limited to whatever the student pastes into the conversation window, and even where structured context is explicitly provided, LLM-generated tutoring responses have been shown not to adapt to it reliably~\cite{borchers2025can}.

The evaluation landscape compounds this: Maurya et al.\ propose a unified taxonomy for assessing LLM-based tutors precisely because existing frameworks lack the granularity to evaluate pedagogical behaviours beyond surface-level response quality~\cite{maurya2025unifying}. Similarly, no standard methodology exists for evaluating the cross-course generalisation of adaptive tutoring architectures~\cite{latif2026systematic}, a gap the three-course deployment reported here is designed to address. The result is a persistent \emph{deployment--evidence gap}: systems are being placed in classrooms before the field has developed the frameworks to evaluate whether they work, and before purpose-built adaptive ITS have been empirically tested against the realities that those classrooms impose~\cite{mannekote2024large, maurya2025unifying}. A related but distinct gap concerns scalability: the capacity of an adaptive tutoring architecture to generalise across disciplinary domains without modification is widely claimed in the ITS literature, but rarely empirically demonstrated~\cite{latif2026systematic, dong2025build}.

Our prior work, \textit{LEA}, addressed the ITS architecture gap~\cite{rumble2026learning}. It is an agentic AI tutoring system that delivers adaptive, personalised instruction through three integrated modes (Chat, Tutor, and Quiz) grounded in a course-specific RAG knowledge base and a hierarchical KC model. Its agent orchestrator continuously estimates the learner's CL, ZPD, and motivational state to dynamically adjust scaffolding type, task difficulty, and feedback timing. Evaluated through simulation-based testing with 14,143 synthetic interactions across six learner profiles, \textit{LEA} demonstrated robust retrieval accuracy, coherent multi-turn tutoring, and adaptive stability~\cite{rumble2026learning}. Simulation, however, does not address the deployment-evidence gap entirely: \emph{does \textit{LEA} work with real students in a real classroom}, and \emph{does it scale across disciplinary domains beyond the graduate-level STEM course for which it was originally designed?}

The central research questions driving this extended work are: (RQ1) How do real students experience \textit{LEA} across its three instructional modes, and where does classroom behaviour diverge from simulation predictions? (RQ2) Can \textit{LEA}'s course-agnostic architecture be operationalised across courses of differing academic levels and disciplinary domains without architectural modification, and if so, what aspects of the system hold unchanged versus require course-specific adaptation?

To answer these questions, we report three new contributions that substantially extend the conference paper. First, we describe \textit{LEA}'s production deployment across three university courses at Abertay University, Dundee: CMP511 (Machine Learning and Artificial Intelligence, MSc), CMP202 (Data Structures and Algorithms 2, undergraduate, UG), and PSY555 (Human Psychology: In what ways are we all the same?, MSc). This three-course deployment spans two degree levels and two disciplinary domains, providing the first empirical account of \textit{LEA}'s scalability claims. The inclusion of PSY555, a non-computing social-science course with discursive, non-procedural content, constitutes a cross-domain stress test. The central architectural scalability claim is that \emph{the orchestration layer generalises; the knowledge layer adapts}. Second, we report the first pilot classroom study of \textit{LEA}, conducted with students enrolled in CMP511 under ethics approval and pseudonymised data collection, using a nine-section survey instrument to assess usability, mode-specific perceptions, learning outcomes, trust, and technical reliability. Third, we report a scalability evaluation (660 questions) across all three courses using the Retrieval Augmented Generation Assessment (RAGAS) framework to measure retrieval and generation quality.

These contributions advance the agenda called for by Mannekote et al.~\cite{mannekote2024large}: moving beyond prompt-response interactions and synthetic evaluation toward systems that are interpretable, pedagogically grounded, and validated against the complex ecology of real learners. The remainder of this paper is structured as follows. Section~\ref{sec:related} reviews related work, expanding the conference paper's background with literature on human-subjects evaluation of AI tutors and cross-course ITS scalability. Section~\ref{sec:background} summarises the psychological and pedagogical foundations underlying \textit{LEA}'s design, including cognitive load theory, Zone of Proximal Development, and motivational frameworks, alongside the RAG and KC modelling principles that ground its content delivery. Section~\ref{sec:architecture} details the system architecture. Section~\ref{sec:scalability} presents the multi-course deployment methodology. Section~\ref{sec:simulation} summarises the simulation-based evaluation reported in the conference paper~\cite{rumble2026learning} and reflects on its limitations, establishing the predictions against which the classroom deployment is subsequently assessed. Section~\ref{sec:human_study} reports on the CMP511 pilot classroom study, including a direct comparison between simulation predictions and observed student experience. Section~\ref{sec:scalability-eval} presents the cross-course scalability evaluation. Section~\ref{sec:discussion} discusses these findings in relation to RQ1 and RQ2, and Section~\ref{sec:conclusion} concludes with a summary, limitations, and directions for future work.

\section{\uppercase{Related Work}}
\label{sec:related}

This section surveys four bodies of prior work that situate \textit{LEA} within the broader landscape of adaptive tutoring research: the evolution of AI-driven tutoring systems, from early cognitive modelling through LLM-driven and agentic architectures (Section~\ref{subsec:its_evolution}); the use of RAG in educational systems (Section~\ref{subsec:rag}); the limited and largely simulation-based evidence base for human-subjects evaluation of AI tutors (Section~\ref{subsec:human_eval}); and cross-course ITS scalability, an architectural property that is frequently claimed in the literature but rarely empirically demonstrated (Section~\ref{subsec:scalability_lit}).

\subsection{From Rule-Based Systems to Agentic AI Tutors}
\label{subsec:its_evolution}

Intelligent tutoring systems have evolved over three broad generations. Early cognitive modelling systems like \textbf{MATHia} (Carnegie Learning) and \textbf{ALEKS} (McGraw-Hill) are examples of learner mastery tracking through Bayesian Knowledge Tracing and Knowledge Space Theory respectively, while \textbf{Squirrel AI} extends this through predictive scaffolding and behavioural analytics~\cite{ritter2007cognitive,cosyn2021practical,Li2025SquirrelAI}. All three achieve strong adaptive assessment from static, pre-authored content libraries, but lack generative capabilities or affective adaptation. A second generation of LLM-driven generative tutors, including \textbf{OATutor}, \textbf{LiveHint AI}, and \textbf{Duolingo AI} offer improved adaptability through fine-tuned language models and GPT-4-based RAG, but have limited learner-state modelling and no cognitive load estimation~\cite{pardos2023oatutor, CarnegieLearning_LiveHintAI_2023, duolingo2023max}.

A third, more recent wave comprises agentic systems with structured course grounding. \textbf{Khanmigo} integrates GPT-4 with Khan Academy's curriculum via RAG, providing strong Socratic scaffolding but no cognitive load monitoring~\cite{shetye2024evaluation}; \textbf{Jill Watson} draws on instructor-supplied materials across seventeen courses but does not provide scaffolding or estimate mastery~\cite{taneja2024jill}; and \textbf{Harvard CS50's AI Tutor} integrates course-specific RAG with careful scaffolding but lacks mastery estimation or a motivational framework~\cite{liu2025improving}. Beyond these dedicated systems, major LLM providers released general-purpose learning interfaces (OpenAI's ChatGPT Study Mode, Google's Gemini Learn LM, and Anthropic's Claude for Education~\cite{openai2025studymode,google2025learnlm,anthropic2025education}) which provide Socratic-style prompting with strong content generation but no KC model capability, mastery estimation, persistent learner-state tracking, or course-specific knowledge grounding. See Table~\ref{tab:its_comparison} for comparison of these ITS platforms.

\begin{table*}[t]
\caption{Comparative analysis of AI tutoring systems, extended to include general-purpose LLM learning features released in 2024--2025 (lower block). Ratings: \textbf{Strong} (S), \textbf{Moderate} (M), \textbf{Basic} (B), \textbf{None} (--).~\cite{openai2025studymode,google2025learnlm,anthropic2025education,shetye2024evaluation,liu2025improving,Li2025SquirrelAI,pardos2023oatutor,CarnegieLearning_LiveHintAI_2023,duolingo2023max,cosyn2021practical,taneja2024jill}}
\label{tab:its_comparison}
\footnotesize
\setlength{\tabcolsep}{4pt}
\begin{tabular}{llccccccc}
\toprule
\textbf{System} &
\textbf{Domain} &
\textbf{KC} &
\textbf{Mastery} &
\textbf{Adaptive} &
\textbf{Course} &
\textbf{Cog.} &
\textbf{Motiv-} &
\textbf{Content} \\
 & & \textbf{Model} & \textbf{Estim.} & \textbf{Scaffold} & \textbf{RAG} & \textbf{Load} & \textbf{ation} & \textbf{Gen.} \\
\midrule
\multicolumn{9}{l}{\textit{Purpose-built adaptive ITS}} \\
\midrule
MATHia          & K-12 Math          & S & S & S & --  & M & M & B \\
ALEKS           & STEM      & S & S & S & --  & -- & M & -- \\
Squirrel AI     & K--12   & S & S & S & --  & M  & M  & B \\
OATutor         & STEM               & S & S & S & --  & -- & M & S \\
LiveHint AI     & Math               & B & B & S & --  & -- & M & S \\
Jill Watson     & Higher Ed  & B & -- & -- & S  & -- & M & M \\
Khanmigo        & Multi-subject      & B & M & S & S  & -- & S & M \\
Harvard CS50    & Comp Science   & S & -- & S & S  & -- & B & S \\
Duolingo AI     & Languages  & B & M & S & M  & -- & S & S \\
\midrule
\multicolumn{9}{l}{\textit{General-purpose LLMs with learning overlays}} \\
\midrule
ChatGPT         & General& -- & -- & B & --  & -- & -- & S \\
Study Mode      & & & & & & & & \\
\addlinespace[2pt]
Gemini & General& -- & -- & M & -- & M & M & S \\
Learn LM        & & & & & & & & \\
\addlinespace[2pt]
Claude for & General& -- & -- & B & --  & -- & -- & S \\
Education            & & & & & & & & \\
\midrule
\multicolumn{9}{l}{\textit{LEA (this work)}} \\
\midrule
\textbf{LEA}    & \textbf{Higher Ed} & \textbf{S} & \textbf{S} & \textbf{S}
                & \textbf{S} & \textbf{M} & \textbf{M} & \textbf{S} \\
                & \textbf{multi-course} & & & & & & & \\
\bottomrule
\end{tabular}
\vspace{2pt}

\noindent\footnotesize
\textit{Notes.} KC Model = structured Knowledge Component hierarchy; Mastery Estim. = probabilistic or threshold-based mastery tracking; Adaptive Scaffold = dynamic scaffolding adjusted to learner state; Course RAG = course-specific RAG; Cog. Load = cognitive load monitoring; Motivation = motivation state monitoring; Content Gen. = LLM-based content generation. \textit{LEA}'s ratings reflect the architecture and simulation evaluation of the conference paper~\cite{rumble2026learning} and the deployments reported here. Ratings for the general-purpose LLM block reflect product documentation~\cite{openai2025studymode,google2025learnlm,anthropic2025education} rather than independent empirical testing; their ``adaptive'' scaffolding is prompt-level Socratic questioning with no persistent learner-state representation across turns.
\end{table*}

Across all three generations, leading platforms achieve strong mastery tracking \emph{or} generative content quality, but few integrate RAG-enhanced generation with cognitive and motivational modelling in a unified, domain-agnostic framework~\cite{latif2026systematic}. \textit{LEA} addresses this gap through three design principles: (1) continuous learner-state awareness via agents tracking cognitive load, proximal performance, and motivational state; (2) adaptive orchestration of scaffolding, task difficulty, and motivational feedback; and (3) pedagogically grounded content generation, anchored in a course-specific RAG vector database and aligned with learning objectives through KC modelling.

\subsection{RAG in Education}
\label{subsec:rag}

RAG integrates document retrieval with LLM content generation, grounding responses in course-specific materials rather than relying on general model knowledge~\cite{lewis2020retrieval}. By incorporating course knowledge into the generative process, RAG architectures reduce hallucination and help ensure contextual accuracy and curricular alignment~\cite{li2025retrieval}. Recent educational deployments confirm the practical value of this approach: RAG-augmented systems outperform conventional static tutors by grounding feedback in course-specific content~\cite{liu2025lpitutor}; combined prompt engineering, RAG, and fine-tuning improve pedagogical adherence over a baseline LLM in a robotics tutor~\cite{kahl2024evaluating}; and knowledge-graph-enhanced RAG improves domain specificity for structured educational content~\cite{dong2025build}. A consistent finding across this literature is that retrieval quality depends on the chunking strategy and embedding model used during corpus preparation, and that these parameters interact with domain characteristics: cross-course benchmarking shows that identical chunker-embedding configurations yield systematically different retrieval quality across disciplinary domains~\cite{stabler2025impact}, and generic chunking heuristics have been shown to fragment the semantic structure of source code specifically, degrading retrieval in ways that do not affect discursive prose~\cite{zhang2025cast}.

The RAGAS~\cite{es2024ragas} suite of metrics has emerged as a standard evaluation framework for RAG pipeline quality. Its four metrics (Table~\ref{tab:ragas_metrics}), measure faithfulness and answer relevancy reference-free, with context precision and recall computed against a reference answer. \textit{LEA} uses RAGAS as the primary metric for cross-course scalability evaluation (Section~\ref{sec:scalability-eval}), enabling comparison of RAG performance across the deployed courses.

\begin{table*}[!ht]
\caption{The four RAGAS evaluation metrics~\cite{es2024ragas} used in the cross-course scalability evaluation (Section~\ref{sec:scalability-eval}). All metrics are scored in $[0,1]$, higher is better.}
\label{tab:ragas_metrics}
\footnotesize
\setlength{\tabcolsep}{4pt}
\begin{tabular}{L{2.2cm}L{2.4cm}L{5.6cm}C{1.8cm}}
\toprule
\textbf{Metric} & \textbf{What It Measures} & \textbf{How It Is Calculated} & \textbf{Reference Answer Required?} \\
\midrule
Faithfulness & 
Factual consistency of the generated answer with the retrieved context &
The answer is decomposed into individual claims via LLM prompting; each claim is verified against the retrieved context. Score is the proportion of claims supported by the context. &
No \\
\addlinespace[2pt]
Answer Relevancy &
How directly the generated answer addresses the question asked &
An LLM generates candidate questions from the answer; the score is the mean cosine similarity between the embeddings of these generated questions and the original question. &
No \\
\addlinespace[2pt]
Context Precision &
Signal-to-noise ratio of the retrieved context &
Each retrieved context chunk is judged for relevance to the reference answer; the score aggregates the proportion of relevant chunks, weighted by rank position. &
Yes\textsuperscript{a} \\
\addlinespace[2pt]
Context Recall &
Completeness of the retrieved context &
The reference answer is decomposed into sentences; each is checked for attributability to the retrieved context. Score is the proportion of reference sentences the context can support. &
Yes\textsuperscript{a} \\
\bottomrule
\end{tabular}
\vspace{2pt}

\noindent\footnotesize
\textit{\textsuperscript{a}} In the present evaluation, the system's own generated answer is used as a stand-in reference in the absence of human-authored ground truth; these two metrics are therefore interpreted as measures of internal retrieval-generation consistency (Section~\ref{sec:scalability-eval}).
\end{table*}

\subsection{Human-Subjects Evaluation of AI Tutors}
\label{subsec:human_eval}

Despite the proliferation of LLM-based tutoring systems, the empirical evidence base for their classroom effectiveness remains thin. The majority of published evaluations rely on simulation, expert annotation, or automated metrics rather than real learner studies~\cite{latif2026systematic, maurya2025unifying}. Mannekote et al.~\cite{mannekote2025can} make this concern explicit, arguing that existing evaluation approaches, including learner simulation, are insufficient to surface the user-experience, domain-transfer, and trust-related dimensions that emerge only in authentic deployment. Their analysis identifies synthetic-vs-real learner divergence as a fundamental validation challenge and calls for controlled classroom studies.

Where human-subjects studies do exist, they tend to be small-scale, short-term, or focused on a single interaction feature rather than whole-system pedagogical effectiveness. Kahl et al.~\cite{kahl2024evaluating} include a small human evaluation component in their robotics tutor study, assessing helpfulness and trustworthiness alongside automated metrics, but acknowledge the limited power of the resulting comparisons. Deployed LLM‑based tutors have generated usage evidence at greater scale: Liu et al.~\cite{liu2025improving} report classroom telemetry from Harvard’s CS50 AI Tutor across a full course cohort, and Taneja et al.~\cite{taneja2024jill} describe Jill Watson’s deployment across seventeen courses. However, neither is framed as a controlled evaluation of mode-specific pedagogical effectiveness, trust, or learning outcomes against a defined instrument. Maurya et al.~\cite{maurya2025unifying} propose a structured evaluation taxonomy for LLM-based AI tutors that covers pedagogy, engagement, and safety dimensions, but note that existing systems have not been systematically evaluated against this taxonomy in classroom settings. The most consequential evidence currently available is the Wu et al.\ meta-analysis~\cite{wu2026chatgpt}, which synthesises 35 experimental studies (4,193 participants) and reports a moderate positive effect of ChatGPT on learning outcomes (Hedges' $g = 0.670$). However, nearly all constituent studies use ChatGPT as a supplementary resource rather than as a structured adaptive tutor, and the meta-analysis explicitly flags instructional mode as a significant moderator, suggesting that the effect size is meaningfully different when AI is used as a structured tutoring system versus as a general-purpose assistant.

This gap between deployment speed and classroom evidence is the main motivation for the CMP511 pilot study reported in Section~\ref{sec:human_study}. It is, to our knowledge, the first reported human-subjects study of a complete tri-modal adaptive tutoring system spanning usability, mode-specific pedagogical perception, learning outcomes, and AI trust in a higher education course.

\subsection{Cross-Course ITS Scalability}
\label{subsec:scalability_lit}

Cross-course scalability (the capacity of an ITS to be deployed across disciplines without architectural modification) is a widely stated design goal but is rarely empirically demonstrated \cite{dong2025build, latif2026systematic}. Most ITS evaluations are conducted within a single domain, and the knowledge representation choices made in that domain (e.g.\ Bayesian Knowledge Tracing optimised for procedural mathematics, or rule-based ontologies designed for physics) rarely transfer without substantive re-engineering~\cite{latif2026systematic}. Work on cross-course knowledge tracing and transferable learner-state representations has examined whether the same underlying model can serve different content domains~\cite{kaser2024simulated}, but system-level deployment transfer, in which a complete adaptive tutoring architecture is operationalised across disciplines without modification, is a distinct question that has not been addressed with matching rigour. The proliferation of LLM-based systems has altered this landscape: general-purpose language models are \emph{domain-agnostic} but remain course-blind without structured knowledge grounding.

Recent RAG-based architectures have advanced domain-agnostic pipeline designs, but an evidentiary gap remains. Dong et al.~\cite{dong2025build} claim cross-disciplinary scalability as an inherent property of modular knowledge-graph-augmented RAG pipelines, while acknowledging that empirical deployment evidence across disciplines is still lacking. Liu et al.~\cite{liu2025lpitutor} demonstrate successful RAG deployment in computer science and engineering contexts but do not extend evaluation to social science courses. A 2026 systematic review of 86 tutoring-system studies found the same imbalance in literature as a whole, with only 9\% of studies addressing social-science content against 52\% in STEM domains~\cite{latif2026systematic}. STEM and social science disciplines differ not only in content vocabulary and density, but, as the present deployment demonstrates (Section~\ref{sec:scalability}), in the structure of their knowledge components and the granularity of learning objectives, the nature of assessment (procedural versus conceptual), and the chunking characteristics of source material (code and mathematics versus discursive text).

\textit{LEA}'s three-course deployment, detailed in Section~\ref{sec:scalability}, is positioned to address this gap directly: rather than asserting cross-course scalability as an architectural property, it tests the claim empirically across two academic levels and two disciplinary domains, including a social science course with no code-based learning objectives.

\section{\uppercase{Background}}
\label{sec:background}
The adaptive mechanisms described in Section~\ref{sec:architecture} are grounded in four established frameworks from cognitive science and educational psychology, summarised here. Effective ITS design integrates cognitive and motivational principles of learning with course-specific content grounded in course curricula~\cite{vanlehn2011relative,villegas2025adaptive}. Within \textit{LEA}, psychological theories inform how the system interprets learner behaviours and delivers scaffolding, while structured knowledge representations enable precise retrieval and alignment with instructional goals.

\subsection{Psychological Foundations of Learning}
These four frameworks --- Cognitive Load Theory, Zone of Proximal Development and constructivist learning theory, engagement theory, and motivation and self-efficacy theory --- define the Integrated Psychological Framework that informs \textit{LEA}'s learner-state modelling and scaffolding decisions via the Agent Orchestrator (Section~\ref{sec:orchestrator}).

\textbf{Cognitive Load Theory (CLT)} holds that working memory is limited and that instruction must balance intrinsic, extraneous, and germane load to support schema acquisition in long-term memory \cite{sweller2019cognitive,paas2003cognitive}. Adaptive tutors can infer cognitive effort non-invasively from indicators such as task completion time and accuracy, adjusting pacing and complexity to avoid overload \cite{paas2016cognitive}.
 
\textbf{Zone of Proximal Development (ZPD) and constructivist theory} hold that learning is most effective within the band between what a learner can do independently and what they can achieve with guidance \cite{wood1976role,vygotsky1978mind}, and that active strategies such as elaborative interrogation and problem-based learning support deeper cognitive construction \cite{freeman2014active}. Adaptive systems operationalise this by tracking knowledge state and fading scaffolding as competence increases \cite{hariyanto2025artificial}.
 
\textbf{Engagement Theory}, grounded in Self-Determination Theory (SDT) and Situated Expectancy-Value Theory, holds that autonomy, competence, and relatedness sustain participation \cite{ryan2000self}, with interface design (mode choice, progress visualisation, collaborative features) directly supporting these needs \cite{peters2018designing,lee2025awareness}.
 
\textbf{Motivation and Self-Efficacy Theory}, drawing on Bandura's Social Cognitive Theory and Zimmerman's model of self-regulation \cite{bandura1997self,zimmerman2002becoming}, influences \textit{LEA}'s use of progressive scaffolding, vicarious modelling, and personalised feedback to build learner confidence and persistence.
 
\subsection{Retrieval-Augmented Generation (RAG) Library}
The \textit{LEA} RAG pipeline converts instructional resources, such as lecture notes, slides, and assessments, into a searchable, vectorised knowledge space. The process begins with document ingestion and validation, followed by semantic chunking that segments content into coherent units optimised for LLM context windows. These chunks are embedded using dense neural models to generate high-dimensional semantic vectors. The resulting embeddings are stored in a scalable vector database to support real-time similarity-based retrieval \cite{lewis2020retrieval}. When a learner query is issued, relevant segments are retrieved and reranked before being supplied to the LLM, grounding generated explanations in validated course materials. This integration allows \textit{LEA} to provide consistent, contextually relevant, and pedagogically aligned responses while maintaining adaptability across multiple courses \cite{li2025retrieval}.

\subsection{Knowledge Component (KC) Model}
The KC Model defines the conceptual structure of a course by hierarchically mapping weekly Learning Objectives (LO) to Granular Objectives (GO) \cite{corbett1994knowledge}. \textit{LEA} defines a methodology for educators to specify high-level LOs, which are expanded into aligned GO Knowledge Components using AI-assisted decomposition and metadata tagging. The resulting KC schema provides an outline for \textit{LEA}'s Tutor and Quiz modalities. This hybrid human-AI pipeline combines educator intent with scalable automation, providing a reproducible method for authoring course-specific tutoring environments utilising pre-existing course materials. The RAG Library and KC Model constitute the data foundation that inform and guide \textit{LEA}'s adaptive learner-aware instruction.

\section{\uppercase{System Architecture}}
\label{sec:architecture}
\begin{figure*}[t]
  \centering
  \includegraphics[width=0.9\textwidth]{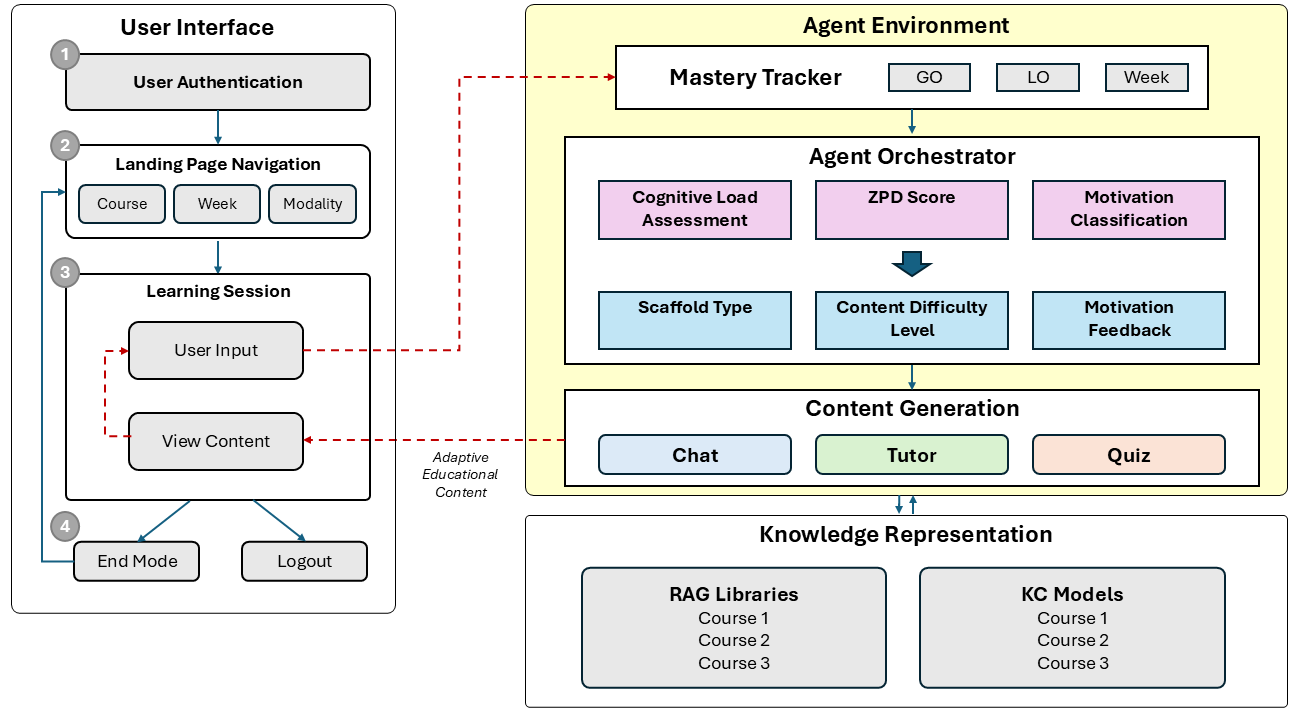}
  \caption{{\textit{LEA}} Agentic Architecture showing the interaction between the User Interface, Agent Environment, and Knowledge Representation layers. Reprinted from our previous conference paper.~\cite{rumble2026learning}}
  \label{fig:agentic_env}
\end{figure*}

\textit{LEA} operates through a distributed, agentic AI architecture designed to deliver adaptive, student-specific instruction by integrating psychological, pedagogical, and computational components into a unified system. The architecture described in Figure~\ref{fig:agentic_env} comprises three core subsystems: \textit{Agent Orchestrator}, \textit{Mastery Tracker}, and \textit{Tri-Modal Content Generation} that collectively enable real-time adaptation while remaining contextually grounded in instructional intent.

\subsection{Agent Orchestrator}
\label{sec:orchestrator}
At the core of \textit{LEA} is an agentic AI environment where multiple specialised agents collaborate to maintain real-time awareness of learner state and adapt instruction accordingly. The Agent Orchestrator performs continuous assessment of CL, ZPD, and motivational state using non-invasive behavioural and performance indicators~\cite{paas2003cognitive,bjork2011making,kochmar2020automated}. These indices drive scaffolding decisions and dynamic task difficulty adjustments.

\subsubsection{Cognitive Load and ZPD Assessment}
Cognitive load is calculated as a weighted combination of task difficulty, recent accuracy, and fatigue indicators (Eq.\eqref{eq:cognitive_load}), enabling detection of overload or underload conditions. The system interprets CL values on a 1–9 Paas rating scale~\cite{paas2003cognitive}, to determine scaffolding intensity and fading rate. The coefficients were determined via offline Monte Carlo simulations (calibration remains open to refinement, Section~\ref{sec:conclusion}).

\begin{equation}
\label{eq:cognitive_load}
CL = 0.5 + 1.0 G + 4.5 (1 - A) + 0.1 Q + 1.0 T_{\text{type}}
\end{equation}

\noindent
where:
\begin{itemize}
    \item $G$ : normalized difficulty of the next GO
    \item $A$ : mean accuracy across the last four responses
    \item $Q$ : number of continuous learner interactions
    \item $T_{\text{type}}$ : next task type (1 = open-ended, 0 = all other)
\end{itemize}

The learner's ZPD is continuously estimated as the mean accuracy across the last four responses (Eq.\eqref{eq:zpd_score}), targeting a performance band of 50--80\%, a design choice informed by desirable-difficulty research~\cite{bjork2011making}. 

\begin{equation}
\label{eq:zpd_score}
ZPD = \frac{1}{n} \sum_{i=t-n+1}^{t} x_i
\end{equation}

\noindent
where:
\begin{itemize}
    \item $ZPD$ : zone of proximal development score
    \item $n$ : window size (4 most recent responses)
    \item $t$ : current time or interaction index
    \item $x_i$ : correctness of response $i$ ($1$ is correct)
\end{itemize}

\begin{figure}[ht]
  \centering
  \includegraphics[width=0.9\linewidth]{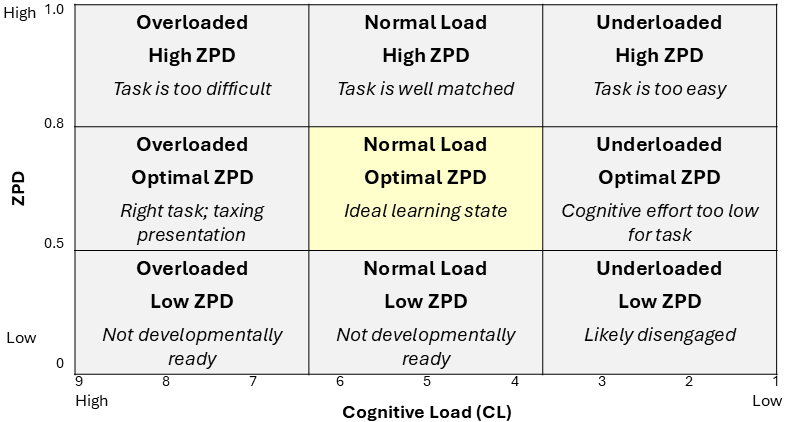}
  \caption{Cognitive Load $\times$ ZPD Scaffolding Decision Matrix guiding adaptive scaffolding logic. Reprinted from our previous conference paper.~\cite{rumble2026learning}}
  \label{fig:cl_zpd}
\end{figure}

These measures enable \textit{LEA} to maintain challenge levels aligned with the learner’s proximal capability. Figure~\ref{fig:cl_zpd} demonstrates how CL and ZPD are used to guide scaffolding choices and to calibrate task difficulty.

\subsubsection{Motivation State Classification}

Motivation state is inferred non-invasively from persistence, sentiment, and performance trends (Eq.\eqref{eq:motivation_change}), following work showing ITS can infer learner state from behavioural data without self‑report~\cite{kochmar2020automated}. We define four states aligned with the tri‑modal interaction structure: \emph{cold start}, when insufficient history prevents classification~\cite{pian2020gamified}; \emph{motivation drop}, reflecting declines in $\Delta P$, $\Delta S$, $\Delta A$; \emph{plateau}, indicating stable but non-improving state; and \emph{maintained high motivation}, indicating sustained positive trends across all three signals. Each state triggers a corresponding instructional strategy grounded in Self-Determination Theory~\cite{ryan2000self}: autonomy support during cold start, competence celebration and affective intervention during motivation drop, challenge escalation during plateau, and adaptive fading during maintained high motivation. Motivational feedback is integrated contextually within the Tutor and Quiz dialogues to sustain engagement and self-efficacy.
\begin{equation}
\label{eq:motivation_change}
\Delta M = \Delta P + \Delta S + \Delta A
\end{equation}

\noindent
where:
\begin{itemize}
    \item $\Delta M$: change in overall motivation level
    \item $\Delta P$: change in persistence (session continuation or interaction frequency)
    \item $\Delta S$: change in sentiment (derived from affective cues in learner responses)
    \item $\Delta A$: change in answer accuracy (reflecting changes in learner competence)
\end{itemize}

\noindent
Positive $\Delta M$ values indicate rising motivation and engagement, prompting adaptive fading or autonomy reinforcement, whereas negative shifts trigger supportive scaffolding or affective interventions to restore learner momentum.

\subsubsection{Scaffolding and Feedback Loop}
The Agent Orchestrator integrates these cognitive and motivational indices to select scaffolding type (conceptual, procedural, strategic, or meta-cognitive) and support intensity \cite{hannafin2013open}. Fading logic reduces scaffolding once stable performance or high motivation is detected. All orchestration events are logged, supporting transparent evaluation and continuous optimisation of adaptive strategies.

\subsection{Mastery Tracking}
The mastery tracking system monitors learner progress at three hierarchical levels, Course Week, Learning Objective (LO), and Granular Objective (GO), storing dynamic mastery indicators alongside each learner profile. The system employs differing assessment pathways: for quiz interactions, it uses direct scoring with complexity-adjusted mastery calculations. For tutor interactions, \textit{LEA} uses an LLM for response evaluation in an adapted Difficulty-aware Programming Knowledge Tracing (DPKT) \cite{yang2025difficulty}, supporting real time, multi-level assessment.

\subsection{Tri-Modal Content Generation}
\textit{LEA} supports three instructional modalities—Chat, Tutor, and Quiz—each grounded in course-specific RAG content. The Chat mode facilitates open exploration of the domain knowledge base and Q\&A interactions. Tutor mode provides structured, Socratic dialog guided by adaptive scaffolding and motivational feedback. Quiz mode offers formative assessment through adaptive question generation and performance evaluation. The Tutor and Quiz modes feed student state data to the Mastery Tracker and Orchestrator, ensuring continuous adaptation across learning sessions.

\section{\uppercase{Multi-Course Scalability: Deployment Methodology}}
\label{sec:scalability}

\subsection{Course-Agnostic Design Principle}
\label{sec:scalability-design}

\textit{LEA}'s key architectural benefit is that scalability does not require redesigning the adaptive engine for each new course. The system achieves this through a strict separation of concerns between \textit{orchestration logic} and \textit{knowledge representation}. The Agent Orchestrator, Mastery Tracker, and scaffolding/motivation engines operate entirely on abstract KC identifiers, cognitive state variables, and learner interaction records, none of which encode course-specific content. Course-specific knowledge is instead confined to two interchangeable artifacts: the RAG library (a ChromaDB vector store of course materials) and the KC model (a hierarchical JSON representation of weekly Learning Objectives and Granular Objectives). Integrating a new course therefore consists of populating these two artifacts rather than modifying the orchestration code base, consistent with the course-agnostic scalability principle articulated in the conference paper \cite{rumble2026learning} and illustrated in Figure~\ref{fig:agentic_env}.

This design was tested through deployment across three courses spanning two disciplines and two academic levels (Table~\ref{tab:course-profile}); each required only course-specific RAG and KC artifacts, with no changes to the orchestrator, mastery tracker, or scaffolding engine. The single exception, a course-blind code-question generation routine within the content-generation layer, is documented in Section~\ref{sec:challenges}.

\subsection{Course Profiles}
\label{sec:course-profiles}

Table~\ref{tab:course-profile} summarises the three deployments discussed in this section.

\begin{table*}[!ht]
\caption{Comparative course profile across the three \textit{LEA} deployments.}
\label{tab:course-profile}
\centering
\begin{tabular}{lccc}
\hline
\textbf{Dimension} & \textbf{CMP511} & \textbf{CMP202} & \textbf{PSY555} \\
\hline
Course & Machine Learning & Data Structures & Human Psyc. \\
Domain & Computing & Computing & Social Science \\
Level & MSc & UG & MSc \\
Weeks & 11 & 10 & 13 \\
Learning Objectives (LOs) & 48 & 30 & 16 \\
Granular Objectives (GOs) & 161 & 90 & 79 \\
Code-based GOs\textsuperscript{a} & 15 (9.3\%) & 32 (35.6\%) & 0 (0\%) \\
RAG document count & 90 & 112 & 64 \\
RAG chunk count & 1{,}266 & 1{,}230 & 691 \\
\hline
\end{tabular}

\textit{\textsuperscript{a}} Code-based GOs are those tagged with a procedural skill category (KC model).
\end{table*}

\textbf{CMP511: Machine Learning and Artificial Intelligence (MSc).} CMP511 was the original deployment target and baseline against which CMP202 and PSY555 are compared. It is a postgraduate, technical STEM course covering supervised, unsupervised, and reinforcement learning, combining conceptual material with Python-based implementation exercises. CMP511's KC model and RAG library were developed iteratively over the course of the original system build and were the basis for the simulation-based validation reported in Section~\ref{sec:simulation} and in the conference paper \cite{rumble2026learning}.

\textbf{CMP202: Data Structures and Algorithms 2 (UG).} CMP202 is an undergraduate course in the same disciplinary area as CMP511 (computing) but at a lower academic level and with a distinct technical focus. Building on its prerequisite, it introduces parallel programming on shared-memory and GPU architectures and the design techniques underpinning parallel applications, illustrated through case studies drawn from real-world workloads. This deployment was selected to test \emph{level scalability}: whether the architecture generalises to an undergraduate cohort. CMP202's code-based content required the system's code-question generation logic to target C\nolinebreak{}++ rather than the Python used in the CMP511 implementation, a distinction discussed further in Section~\ref{sec:challenges}.

\textbf{PSY555: Human Psychology (MSc).} PSY555 is a postgraduate course in cognitive human psychology, covering topics including neural architecture and perception, memory models, attention, and decision-making. It contains no programming or code-based content. PSY555 was selected as the cross-domain test case: a social-science course with discursive learning objectives. PSY555's KC model contains zero procedural Granular Objectives, making it a direct test of whether the KC model schema, and the system's downstream content-generation logic, can accommodate domains with no code-based learning objectives.

\subsection{Course Onboarding Pipeline}
\label{sec:onboarding-pipeline}

Adding a new course to \textit{LEA} follows a fixed three-stage pipeline, consistent across all three courses deployed in this study.

\textbf{Stage 1: RAG library creation.} Course materials (lecture slides, lecture recordings, tutorial notes, and, where applicable, code files) are placed into a per-course document repository following a standardised naming convention. An automated pipeline performs file validation and change detection via MD5 signature comparison, format-specific conversion to Markdown (PDFMiner for PDFs, Whisper for video transcription, MarkItDown for Office documents), GPT-4-based document analysis and summarisation, semantic chunking (500-token windows with 100-token overlap, sentence-bounded via NLTK), and embedding generation using OpenAI's \texttt{text-embedding-3-small} model. Embeddings are persisted to a dedicated ChromaDB collection, preventing cross-contamination between course knowledge bases.

\textbf{Stage 2: KC model authoring.} An instructor completes a structured Excel template specifying, for each week, a description and a list of Learning Objectives with references to supporting RAG materials. An automated extraction script decomposes each Learning Objective into Granular Objectives, assigns Bloom's-taxonomy cognitive level classifications, sets mastery thresholds, and identifies cross-week dependencies, producing an Excel output for review.

\textbf{Stage 3: Human-in-the-loop review.} The instructor reviews and edits the generated KC structure; accepting, editing, adding, or deleting Granular Objectives as needed, before the reviewed spreadsheet is converted to the final, KC Model file.

This pipeline was applied identically across CMP511, CMP202, and PSY555, with the only variation being the content and structure of the source materials and instructor-authored learning objectives themselves.

\subsection{A Course-Blind Failure Mode: Code-Question Generation}
\label{sec:challenges}
One component, however, was not course-agnostic in practice. Quiz mode's logic hard-coded Python question generation as a default (a choice originally made because Python fluency is a core objective of CMP511) and relied on generic Granular Objective metadata with no awareness of which course was active. As a result, the system generated Python code questions for PSY555, a course with no programming content. For CMP202, the same logic generated Python rather than the course's actual language (C\nolinebreak{}++), since the underlying example code throughout the system had only been written with CMP511 in mind. This was resolved by adding an explicit mapping from course to programming language, so that PSY555 never generates code questions and CMP202 generates C\nolinebreak{}++ rather than Python questions. This was the only cross-course failure requiring a code change during onboarding; all other adaptations were confined to RAG library and KC model artefact generation.

\section{\uppercase{Simulation-Based Evaluation}}
\label{sec:simulation}

\textit{LEA} was previously assessed using simulation-based testing with synthetic learner agents, consistent with established ITS evaluation practice~\cite{matsuda2007evaluating}. The full simulation methodology and results are reported in the conference paper~\cite{rumble2026learning}\footnote{Source code and illustrative datasets are available at https://github.com/tep00018/LEA.}; they are summarised below to establish the predictions against which the classroom deployment is currently assessed. Six learner profiles grounded in established learner characteristics (prior knowledge, cognitive load tolerance, motivation, engagement style, and learning rate~\cite{kaser2024simulated}) were constructed following the approach of profiling synthetic students against the course's Knowledge Components~\cite{lu2024generative}. These profiles interfaced with \textit{LEA}'s full architecture across three studies: Chat ($n=1{,}980$), Tutor ($n=5{,}563$ turns), and Quiz ($n=6{,}600$ items), with each generative response scored against course content using an LLM-as-a-Judge framework~\cite{zheng2023judging}.

Chat Mode achieved 100\% retrieval success and $M=0.880$ generated answer quality (95\% CI [0.879, 0.882]), both exceeding target. Tutor Mode exceeded target on both Multi-Turn Effectiveness ($M=0.861$, 95\% CI [0.859, 0.863]) and Adaptive Feedback Appropriateness ($M=0.807$, 95\% CI [0.805, 0.809]). A two-way analysis of variance (ANOVA) on the \textit{Learner State Match} metric found that learner profile, not instructional week, was the dominant source of variation ($F(5,5495)=26.70$, $p<.001$, $\eta^2_p=0.018$ for profile vs.\ $\eta^2_p=0.002$ for week), indicating high temporal stability across the simulated term. Quiz Mode's KC Alignment Score averaged $M=0.751$ (95\% CI [0.745, 0.757]), below target, with thematic analysis~\cite{braun2006using} of feedback logs suggesting prompt phrasing favoured recall over higher-order reasoning. Aggregate response accuracy ($M=0.398$) varied strongly by learner profile ($F(5,6585)=107.68$, $p<.001$, $\eta^2_p=0.974$), confirming sensitivity to learner capability.

\textbf{Reflection on Simulation Limitations.} This design is well suited to questions of internal consistency (i.e. whether the orchestrator responds differently to different synthetic learner states), but is less well suited to three classes of question that only the subsequent classroom deployment (Section~\ref{sec:human_study}) and cross-course analysis (Section~\ref{sec:scalability-eval}) could address:

\begin{itemize}
    \item \textbf{User experience.} Synthetic agents respond to scaffolding and feedback according to parameterised rules; they do not experience Tutor Mode's Socratic loop as rigid, or its motivational affirmations as performative, in the way that two CMP511 participants independently reported (Section~\ref{sec:human_study}). The simulation framework measures whether scaffolding was \emph{appropriate} by an external rubric, not whether it \emph{felt} appropriate to the person receiving it.
    
    \item \textbf{Real cognitive load.} The orchestrator's CL and motivation estimates were validated against synthetic learners whose internal states were generated by the same family of formulae the orchestrator uses to infer them, a circularity no purely synthetic evaluation can escape. The classroom pilot's qualitative reports of scaffolding rigidity and affirmation fatigue (Section~\ref{sec:human_study}) are the first evidence of how these models perform against learners whose states they did not generate.

    \item \textbf{Domain transfer.} All simulation-based evaluation reported above was conducted exclusively against CMP511 content, prior to \textit{LEA}'s deployment to CMP202 or PSY555. The Faithfulness gradient identified in Section~\ref{sec:scalability-eval} (0.685 for CMP511 to 0.499 for PSY555) is a finding that simulation could not have reproduced, since the evaluation was cross-week within CMP511 only and not cross-course.
    
\end{itemize}

These are limits on what synthetic-only evaluation on a single domain can establish, not flaws in the simulation methodology. Section~\ref{sec:human_study} and Section~\ref{sec:scalability-eval} test these boundaries and report the divergences.

\section{\uppercase{Exploratory Pilot Classroom Study}}
\label{sec:human_study}

\subsection{Study Design, Instrument, and Participants}
\label{sec:ethics}

The CMP511 pilot constitutes the first deployment of \textit{LEA} in a live classroom context. Following simulation-based validation~\cite{rumble2026learning}, the study aimed to characterise student experience, perceived usability, and mode-specific pedagogical utility, and to provide an empirical basis for comparing simulation predictions against real-world behaviour. Ethical approval was granted by the Abertay University Research Ethics Committee (ref.\ EMS11930, 9 March 2026) covering informed consent, voluntary participation, right of withdrawal, GDPR (General Data Protection Regulation)-compliant data handling, and pseudonymisation. 

Participants were students enrolled in CMP511: Machine Learning and AI (MSc, $N=41$), 2025--26. \textit{LEA} was available to the full cohort as a voluntary study aid via pseudonymised login. A nine-section survey (7-point Likert scale; Sections A--I covering background, usability, three mode experiences, learning outcomes, trust, technical issues, and open design preferences) was distributed towards the end of term, with several items reversed to detect inattentive responding. The full instrument, including exact item wording and all reversed items, is provided as Supplementary Information (Online Resource~1). Eight responses were received (20\% response rate); the study is treated throughout as an \textit{exploratory pilot}. No inferential statistics are reported on Likert data; quantitative items are summarised descriptively using frequency distributions and the proportion rating $\geq$5/7. Open-text responses are analysed using Braun and Clarke's~\cite{braun2006using} reflexive thematic analysis.

All eight participants had at least occasional experience with general-purpose AI tools, though four identified \textit{LEA} as their first dedicated AI tutoring platform. Session engagement was low overall: five participants reported one or two sessions across the term, two reported three to five, and one six to ten —- a shallow engagement depth that is acknowledged as a constraint throughout the findings.

\FloatBarrier
\subsection{Findings}
\label{sec:pilot-findings}

Table~\ref{tab:consolidated_findings} consolidates quantitative ratings across all survey sections ($n=8$; 7-point Likert). Full agreement (8/8, $\geq$5/7) was achieved on all usability items (median~6), confirming a coherent positive usability theme corroborated in open text: participants described the interface as ``simple to use'' and ``clear,'' and independently praised the ability to move between modes within a session.

\begin{table*}[!t]
\centering
\caption{Consolidated pilot study ratings across all survey sections
($n = 8$; 7-point Likert).}
\label{tab:consolidated_findings}
\resizebox{\textwidth}{!}{%
\begin{tabular}{llcc}
\hline
\textbf{Section} & \textbf{Item} & \textbf{Median} & \textbf{$\geq$5/7} \\
\hline
\multicolumn{4}{l}{\textit{B: Usability}} \\
& Easy to navigate and use & 6 & 8/8 (100\%) \\
& Interface clear and well-organised & 6 & 8/8 (100\%) \\
& Comfortable alongside study routine & 6 & 8/8 (100\%) \\
& Command Center\textsuperscript{a} kept focus on right content & 6 & 7/8 (88\%) \\
& Would recommend to CMP511 peers & 6 & 7/8 (88\%) \\
& Overall satisfaction & 6 & 8/8 (100\%) \\
\hline
\multicolumn{4}{l}{\textit{C: Chat Mode}} \\
& Accurate, course-relevant answers & 6 & 8/8 (100\%) \\
& Responses easy to understand & 6 & 8/8 (100\%) \\
& Helped explore topics I was uncertain about & 6 & 8/8 (100\%) \\
& Did not trust information (rev.) & 3.5 & 3/8 (38\%) \\
& Useful complement to lecture material & 6 & 8/8 (100\%) \\
\hline
\multicolumn{4}{l}{\textit{D: Tutor Mode}} \\
& Felt genuinely different from a standard chatbot & 6 & 6/8 (75\%) \\
& Socratic style helped me think more deeply & 6 & 7/8 (88\%) \\
& Adapted appropriately when I said I was unsure & 4 & 3/8 (38\%) \\
& Helped improve understanding of difficult topics & 6 & 6/8 (75\%) \\
& Motivational feedback kept me engaged & 6 & 6/8 (75\%) \\
\hline
\multicolumn{4}{l}{\textit{E: Quiz Mode}} \\
& Questions clearly worded and well-structured & 6 & 8/8 (100\%) \\
& Questions well-aligned with CMP511 content & 6 & 8/8 (100\%) \\
& Difficulty of questions felt appropriate & 6 & 8/8 (100\%) \\
& Feedback helped me understand mistakes & 6 & 8/8 (100\%) \\
& Quiz Mode useful for testing knowledge & 6.5 & 8/8 (100\%) \\
& Progress tracking helped motivation & 5.5 & 7/8 (88\%) \\
\hline
\multicolumn{4}{l}{\textit{F: Perceived Learning Outcomes}} \\
& Helped me better understand CMP511 concepts & 6 & 8/8 (100\%) \\
& Felt more confident after using \textit{LEA} & 5.5 & 8/8 (100\%) \\
& Helped identify gaps I might have missed & 6 & 8/8 (100\%) \\
& Adaptive feedback relevant to my learning & 6 & 7/8 (88\%) \\
& Overall engagement with CMP511 was higher & 5.5 & 5/8 (63\%) \\
& Would have learned as much without \textit{LEA} (rev.) & 4 & 3/8 (38\%) \\
\hline
\multicolumn{4}{l}{\textit{G: Trust and AI Attitudes}} \\
& Trust \textit{LEA} to provide accurate information & 6 & 8/8 (100\%) \\
& Comfortable data not personally identifiable & 6 & 8/8 (100\%) \\
& Adaptation encouraged honest engagement & 6 & 7/8 (88\%) \\
& AI tutoring tools belong in university education & 6 & 6/8 (75\%) \\
& Concerned AI replaces human instruction (rev.) & 4 & 3/8 (38\%) \\
& Using \textit{LEA} made me more curious about AI in education & 6 & 6/8 (75\%) \\
\hline
\end{tabular}%
}

\vspace{4pt}
\noindent\footnotesize
\textit{Notes.} Reversed items marked (rev.); for these, $\geq$5/7 indicates agreement with the negative claim.

\textsuperscript{a}\,The Command Center is \textit{LEA}'s landing interface for selecting course, week, and mode (Figure~\ref{fig:agentic_env}).
\end{table*}

\textbf{Chat Mode} was the strongest performer: full agreement (8/8) on accuracy, comprehensibility, usefulness, and lecture complementarity. Two primary use patterns emerged qualitatively: \textit{lecture condensation} (extracting core concepts from a week's material) and \textit{concept clarification} across topics including neural networks, logistic regression, and Random Forests. One participant explicitly compared \textit{LEA}'s Chat responses favourably to ChatGPT and Gemini, attributing the advantage to course-specific grounding, consistent with the design rationale for RAG-grounded ITS over general-purpose alternatives~\cite{mannekote2024large}.

\textbf{Tutor Mode} produced the most divided ratings. The Socratic questioning approach was endorsed by 7/8 participants as helpful for deeper thinking, but only 3/8 agreed that \textit{LEA} adapted appropriately when they expressed uncertainty (median~4, Neutral) —-- the most consequential divergence from simulation predictions in this study (see Section~\ref{sec:simvsreal}). Two participants independently reported \textit{affirmation fatigue}: motivational affirmations felt ``disconnected from the conversation flow,'' with one requesting their removal. A further qualitative theme concerned \textit{scaffolding rigidity}: one participant described copy-pasting \textit{LEA}'s own text back as a response to progress the dialogue, suggesting the Socratic loop did not gracefully handle genuine comprehension failures.

\textbf{Quiz Mode} received uniformly strong structural ratings (8/8 on all five content and difficulty items; progress-tracking, a motivational item, at 7/8), in contrast with the simulation's below-target KC Alignment Score ($M=0.751$, below the 0.80 target), indicating that perceived alignment exceeded the LLM-as-Judge estimate. However, a question-repetition bug was reported by 2/8 participants; one classified it as ``severe'' and reported that it prevented meaningful use. This defect is one that simulation-based evaluation is structurally unable to identify.

\textbf{Learning outcomes} were uniformly positive on understanding (8/8), confidence (8/8), and gap identification (8/8). Engagement uplift was more measured (5/8 agreed; three selected Neutral). The reversed additionality item was endorsed by three respondents; however, these were the same three respondents who endorsed the reversed trust and additionality items while simultaneously affirming the corresponding positively-worded items at equivalent levels —-- an acquiescent response pattern of the kind reversed items are included to detect —-- and notably included the sample's most engaged participant rather than its least.

\textbf{Trust} was unanimous: all eight participants trusted \textit{LEA}'s accuracy and were comfortable with data anonymisation. The apparent tension with Section~C's reversed trust item (3/8 endorsing ``did not trust the information'') is most plausibly read as the acquiescent pattern noted above rather than substantive distrust, although with $n=8$ the two readings cannot be definitively separated. Six of eight agreed AI tutoring tools belong in university education; only three expressed concern about AI replacing human instruction. The dominant qualitative position was complementarity: \textit{LEA} ``is good as a tool to use alongside, but not to replace'' instructor-led teaching.

Five of eight participants reported no technical problems; one experienced the quiz repetition bug, one could not access \textit{LEA} from the university network (attributable to the deployment architecture), and one reported minor latency issues without impact on engagement.

\subsection{Qualitative Thematic Analysis}
\label{sec:themes}

Following Braun and Clarke~\cite{braun2006using}, five themes emerged from inductive analysis of open-text responses across all nine sections:

\begin{itemize}
    \item \textbf{Course-specific grounding as differentiator}: participants consistently preferred \textit{LEA}'s course-anchored responses to general-purpose tools, validating RAG-grounded ITS over unstructured AI~\cite{mannekote2024large}.
    \item \textbf{Structural clarity of the tri-modal design}: fluid movement between Chat, Tutor, and Quiz modes within a session was independently praised.
    \item \textbf{Tutor Mode scaffolding perceived as rigid under genuine confusion}: corroborates, at the experiential level, the Learner State Match weakness identified for low-knowledge profiles in simulation~\cite{rumble2026learning}.
    \item \textbf{Motivational affirmations perceived as performative by some}: SDT-grounded feedback~\cite{ryan2000self} may require calibration for postgraduate learners preferring direct instructional exchanges.
    \item \textbf{\textit{LEA} as complement, not replacement}: participants distinguished individual-level complementarity (welcome) from institutional-level cost substitution (a concern), relevant to the Discussion's ethical considerations.
\end{itemize}

\subsection{Simulation-versus-Reality Comparison}
\label{sec:simvsreal}

The pilot findings above bear directly on how much weight simulation-based validation should carry going forward. Table~\ref{tab:simvreal} presents a mode-by-mode comparison of simulation predictions against human-study observations. Three divergence categories emerge: Chat Mode predictions matched student experience directly; Tutor Mode predictions were overoptimistic in the aggregate while consistent with the simulation's own more granular Learner State Match finding; Quiz Mode predictions were exceeded by student ratings. A fourth pattern, the question-repetition bug, has no simulation analogue. The analytical implications of these divergences are developed in Section~\ref{sec:discussion}.

\begin{table*}[t]
\centering
\caption{Simulation Predictions vs.\ Human Study Observations by Mode}
\label{tab:simvreal}
\resizebox{\textwidth}{!}{%
\begin{tabular}{>{\raggedright}p{1cm}>{\raggedright}p{2.5cm}>{\raggedright}p{2.5cm}>{\raggedright}p{3cm}>{\raggedright}p{2.4cm}}
\hline
\textbf{Mode} & \textbf{Simulation Metric} & \textbf{Simulation Result} &
\textbf{Human Study Observation} & \textbf{Divergence} \\
\hline
Chat & RAG Retrieval Success Rate & 100\% (exceeded 80\% target) &
8/8 rated answers as accurate and course-relevant & Consistent \\
\hline
Chat & Generated Answer Quality (LLM-as-Judge) & $M = 0.880$ (exceeded target) &
8/8 rated responses as easy to understand; perceived favourably vs.\ GPT/Gemini &
Consistent \\
\hline
Tutor & Adaptive Feedback Appropriateness & $M = 0.807$ (met 80\% target) &
Only 3/8 agreed \textit{LEA} adapted appropriately; reports of
scaffolding rigidity & Simulation overestimated adaptive responsiveness \\
\hline
Tutor & Learner State Match (low-knowledge profiles) & Significantly lower scores
for Novice Variable and Struggling Eager profiles &
Participants reported inflexibility in Socratic loop;
affirmation fatigue reported & Simulation finding confirmed at experiential level \\
\hline
Quiz & KC Alignment Score & $M = 0.751$ (below 80\% target) &
8/8 rated questions as well-aligned to CMP511 content &
Simulation underestimated content alignment \\
\hline
Quiz & Technical stability & Not modelled & Question repetition bug; prevented meaningful use for 2/8 participants & Simulation could not anticipate defect \\
\hline
All modes & User trust in system & Not directly modelled & 8/8 trusted accuracy;
data privacy comfort 8/8 & Simulation had no analogue \\
\hline
\end{tabular}%
}
\end{table*}

\subsection{Study Limitations}
\label{sec:study-limitations}

This study has several limitations. The sample ($n=8$, 20\% response rate) precludes inferential analysis and limits generalisability; low session engagement (five of eight participants reporting one or two sessions) means findings reflect initial impressions rather than sustained use. The late-term deployment window may have depressed engagement, and retrospective self-reports introduce standard method limitations. The quiz repetition bug and network access constraint may also have introduced attrition bias. The study should therefore be read as a first-deployment feasibility assessment: it establishes that \textit{LEA} can be deployed in a live classroom and generates initial user-experience evidence, but does not constitute a controlled evaluation of learning efficacy.

\section{\uppercase{Cross-Course Scalability Evaluation}}
\label{sec:scalability-eval}

As established in Section~\ref{sec:simulation}, the simulation-based KC Alignment Score ($M = 0.751$, 95\% CI [0.745, 0.757]) is a single-domain estimate. To test whether retrieval-and-generation quality more broadly holds across domains and academic levels, we evaluated all three deployed courses using the RAGAS framework \cite{es2024ragas} against a course-proportional question set generated directly from each course's KC model. For each Granular Objective, two questions were generated by prompting an LLM with the objective's skill descriptor and week context, yielding questions germane to the stated curriculum independently of whether the RAG corpus contained supporting material. Each question was answered using \textit{LEA}'s live, production Chat Mode retrieval and generation pipeline, with no synthetic learner simulation layer.

Context Precision and Context Recall in the present evaluation use the system's own generated answer as a stand-in reference in the absence of human-authored ground truth, and should be read as measures of internal retrieval-generation consistency rather than alignment with an independent correctness standard (Table~\ref{tab:ragas_metrics}); Faithfulness and Answer Relevancy require no reference and are scored as RAGAS intends. Section~\ref{sec:limitations} discusses this constraint further.

\begin{figure}[t]
\centering
\includegraphics[width=0.9\linewidth]{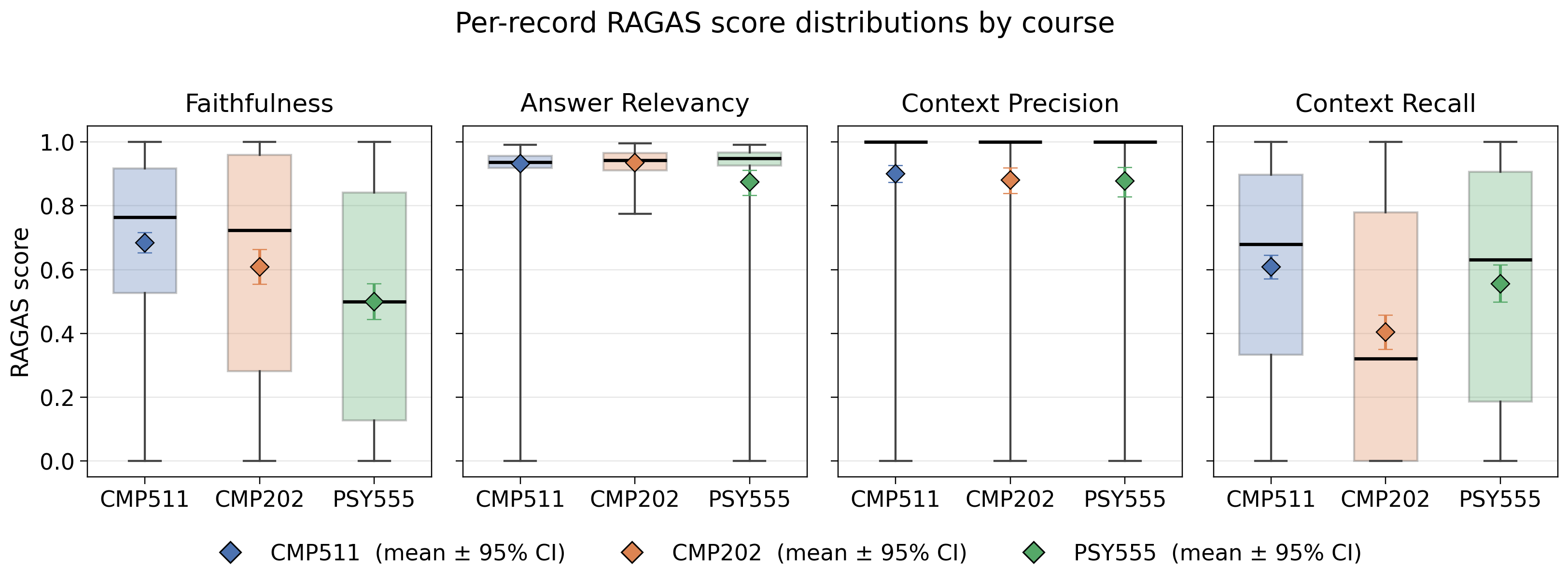}
\caption{RAGAS retrieval and generation quality metrics across the three deployed courses. Faithfulness declines with distance from CMP511's original domain, while Answer Relevancy and Context Precision remain stable across all three courses. Context Recall is notably lower for CMP202 than for either CMP511 or PSY555.}
\label{fig:ragas_chart2}
\end{figure}

Table~\ref{tab:ragas} and Figure~\ref{fig:ragas_chart2} report all four metrics for each course. Answer Relevancy and Context Precision are comparatively stable across domains (0.875--0.935 and 0.877--0.901 respectively), with only a modest decline for PSY555 on Answer Relevancy ($-0.06$ relative to the computing courses), and less than a third of the Faithfulness decline discussed below, indicating that the system consistently retrieves topically relevant material and generates on-topic responses regardless of whether the underlying content is technical STEM material or discursive psychology text. Faithfulness, by contrast, declines across the three courses, from 0.685 (CMP511) through 0.609 (CMP202) to 0.499 (PSY555); as the pairwise tests below establish, this decline is carried by the significant cross-domain drop at PSY555 rather than by the non-significant cross-level step to CMP202. Since the Chat Mode generation prompt was originally authored and tuned exclusively against CMP511 content, this gradient is consistent with the broader pattern identified in Section~\ref{sec:challenges}: auxiliary content-generation logic downstream of the architecturally course-agnostic orchestration layer can encode single-course assumptions that degrade content generation, rather than failing outright, as domain distance increases.

\begin{table*}[!ht]
\centering
\caption{RAGAS retrieval and generation quality across the three deployed
courses. Cell entries are the mean with a bootstrap 95\% confidence interval
($10{,}000$ resamples) beneath. The final column reports the Kruskal--Wallis
omnibus test across courses (distribution-free; scores are non-normal,
Shapiro--Wilk $p<10^{-6}$ throughout). Context Precision and Context Recall
use the system's own generated answer as a stand-in reference in the absence
of human-authored ground truth, and are read as measures of internal
retrieval-generation consistency (Section~\ref{sec:scalability-eval}).}
\label{tab:ragas}
\begin{tabular}{lcccc}
\toprule
\textbf{Metric} & \textbf{CMP511} & \textbf{CMP202} & \textbf{PSY555}
 & \textbf{Kruskal--Wallis} \\
\midrule
Faithfulness
 & 0.685          & 0.609          & 0.499
 & $H=25.01$ \\
 & {\footnotesize [0.652, 0.716]} & {\footnotesize [0.554, 0.663]} & {\footnotesize [0.444, 0.555]}
 & {\footnotesize $p<10^{-5}$, $\varepsilon^2=0.035$} \\
\addlinespace[2pt]
Answer Relevancy
 & 0.933          & 0.935          & 0.875
 & $H=6.15$ \\
 & {\footnotesize [0.925, 0.938]} & {\footnotesize [0.929, 0.941]} & {\footnotesize [0.833, 0.912]}
 & {\footnotesize $p=0.046$, $\varepsilon^2=0.006$} \\
\addlinespace[2pt]
Context Precision
 & 0.901          & 0.881          & 0.877
 & $H=1.11$ \\
 & {\footnotesize [0.874, 0.926]} & {\footnotesize [0.838, 0.918]} & {\footnotesize [0.828, 0.921]}
 & {\footnotesize $p=0.58$ (n.s.)} \\
\addlinespace[2pt]
Context Recall
 & 0.609          & 0.404          & 0.556
 & $H=32.00$ \\
 & {\footnotesize [0.571, 0.645]} & {\footnotesize [0.350, 0.458]} & {\footnotesize [0.498, 0.614]}
 & {\footnotesize $p<10^{-6}$, $\varepsilon^2=0.046$} \\
\midrule
$n$ (questions scored) & 322\textsuperscript{a} & 180 & 158 & \\
\bottomrule
\end{tabular}

\vspace{2pt}
\noindent\footnotesize
\textit{\textsuperscript{a}} One CMP511 record returned a null Faithfulness
score and was excluded from that metric only ($n=321$).
\end{table*}

Because RAGAS scores are bounded on $[0,1]$ and strongly non-normal (Shapiro--Wilk $p<10^{-6}$ for every course--metric combination), cross-course differences were assessed with distribution-free methods: a Kruskal--Wallis omnibus test per metric (Table~\ref{tab:ragas}), pairwise Mann--Whitney $U$ tests with Holm correction and Cliff's $\delta$ effect sizes, and bootstrap 95\% confidence intervals on each mean ($10{,}000$ resamples). The tests sharpen the descriptive pattern. Context Precision does not differ significantly across the three courses, and Answer Relevancy differs only marginally and with a negligible effect size, its sole significant pairwise contrast being CMP511 vs.\ PSY555 ($p_{\text{Holm}}=0.026$, $\delta=-0.15$). Faithfulness differs significantly: PSY555 is significantly lower than both CMP511 ($p_{\text{Holm}}<10^{-6}$, $\delta=0.29$) and CMP202 ($p_{\text{Holm}}=0.005$, $\delta=0.19$), while the CMP511--CMP202 difference is not individually significant ($p_{\text{Holm}}=0.31$); the apparent monotonic decline in mean Faithfulness is therefore driven principally by the cross-domain (PSY555) drop rather than the cross-level (CMP202) one. Context Recall also differs significantly but with a distinct pattern: CMP202 is significantly lower than both CMP511 ($p_{\text{Holm}}<10^{-7}$, $\delta=0.31$) and PSY555 ($p_{\text{Holm}}<10^{-3}$, $\delta=-0.22$), whereas CMP511 and PSY555 do not differ. The recall deficit thus tracks CMP202 specifically rather than disciplinary distance, consistent with the code-density explanation below.

A specific architectural prediction follows from the differing content structures of the three courses: retrieval quality may vary systematically with content type, since CMP511 and CMP202's source materials are code-heavy and notation-dense, while PSY555's are discursive, qualitative text. The RAG pipeline's chunking strategy (Section~\ref{sec:onboarding-pipeline}, Stage~1) was designed and tuned against CMP511 content; its 500-token, sentence-bounded chunking approach was not separately validated against code blocks or against the longer, more elaborated sentence structures typical of psychology source texts.

The RAGAS evaluation provides partial evidence for this prediction. Context Recall is markedly lower for CMP202 (0.404) than for either CMP511 (0.609) or PSY555 (0.556). Given that CMP202's KC model has the highest proportion of procedural Granular Objectives of any of the three courses (32/90, or 35.6\%, against 9.3\% for CMP511 and 0\% for PSY555; Table~\ref{tab:course-profile}), and that its RAG corpus accordingly contains a higher density of code notebooks and lab materials, this recall deficit is consistent with the hypothesis that sentence-bounded chunking degrades retrieval completeness specifically for code-heavy content, plausibly by splitting logically coherent code blocks (functions, cells, or worked examples) across chunk boundaries in ways that scatter relevant material rather than preserving it --- a failure mode independently documented for generic chunking of source code~\cite{zhang2025cast}. We treat this as a plausible explanation rather than a confirmed mechanism; isolating chunking strategy as the causal factor, as opposed to other differences between CMP202's corpus and the other two courses, would require a controlled ablation that was beyond the scope of the present work.

\paragraph{From retrieval metric to curriculum-coverage signal.} A methodological observation emerged from constructing the evaluation set. Because the questions were generated from each course's KC model rather than sampled from the RAG corpus, every question corresponds to a stated learning objective, whether or not the corpus contains material that answers it. Faithfulness, the proportion of generated claims grounded in retrieved context, is consequently depressed not only by generation errors but also by \emph{coverage gaps}: an objective whose supporting content is absent from the corpus yields a correct-but-ungrounded answer that scores low through no fault of the generation pipeline. In the present evaluation this is a confound on the Faithfulness gradient, which we therefore interpret jointly with the single-course generation logic of Section~\ref{sec:RQ2}. It also, however, suggests a constructive use of the same machinery: the divergence between KC-derived and corpus-derived question performance could localise where a course's materials fail to cover its objectives. We note this as a direction for future work (Section~\ref{sec:conclusion}).

\section{\uppercase{Discussion}}
\label{sec:discussion}
This section addresses the two research questions posed in Section~\ref{sec:introduction}, drawing on evidence from the classroom pilot (Section~\ref{sec:human_study}), the simulation-based evaluation (Section~\ref{sec:simulation}), and the cross-course scalability analysis (Section~\ref{sec:scalability-eval}).

\subsection{RQ1: How Do Real Students Experience \textit{LEA}, and Where Does Classroom Behaviour Diverge from Simulation?}

The CMP511 pilot provides the first direct evidence on this question, and the simulation-versus-reality comparison (Table~\ref{tab:simvreal}) is the clearest single answer available. Three distinct patterns emerged, each with a different implication for how simulation-based validation should be weighted against human evidence.

For Chat Mode, simulation was predictively accurate. The 100\% retrieval success rate and $M=0.880$ answer-quality score reported in Section~\ref{sec:simulation} translated directly into student experience: all eight participants rated Chat responses as accurate, course-relevant, and easy to understand, with several explicitly preferring \textit{LEA}'s course-grounded answers to general-purpose tools such as ChatGPT and Gemini. Where simulation predicts a simple, single-turn retrieval-and-generation task performs well, the human data confirms it.

For Tutor Mode, simulation was \emph{partially} overoptimistic. The aggregate Adaptive Feedback score ($M=0.807$) suggested the scaffolding engine was performing above target, but this aggregate obscured a weakness the simulation itself had already flagged at a more granular level: the Learner State Match metric was significantly lower for the Novice\_Variable and Struggling\_Eager profiles (Section~\ref{sec:simulation}). Only three of eight participants agreed that \textit{LEA} adapted appropriately when they expressed uncertainty, and several described the Socratic loop as rigid under genuine confusion. The lesson is not that simulation failed here, but that a single aggregate score can mask a sub-population effect that only becomes visible once real, lower-knowledge learners encounter the system.

For Quiz Mode, simulation was, if anything, conservative. The KC Alignment Score of 0.751 fell below the simulation's own 0.80 target and was treated as a design weakness in the conference paper. Yet all eight pilot participants rated quiz questions as well-aligned to course content. Whether this reflects a calibration gap in the LLM-as-Judge rubric or a genuine difference between objective and perceived alignment remains open, but the direction of the divergence --- human perception exceeding the simulated benchmark --- is the opposite of the Tutor Mode finding, underlining that simulation under- and over-estimates different aspects of the same system rather than failing uniformly in one direction. Simulation also had no mechanism to detect the quiz question-repetition bug that two participants described: an interface defect that only manifests under human interaction.

These findings support treating learner simulation as a necessary but insufficient validation step: it is informative about \emph{response quality} under controlled conditions but blind to \emph{deployment-level} failure modes that only manifest under authentic use, where learner behaviour, attention, and tolerance for interface friction vary in ways synthetic agents do not model.

\subsection{RQ2: Can \textit{LEA}'s Architecture Be Operationalised Across Courses Without Modification?}
\label{sec:RQ2}

The answer is a qualified yes. Across all three deployments, the orchestrator, mastery tracker, and scaffolding logic required zero code changes. However, auxiliary content-generation logic downstream of these layers did require modification to adapt for course-specific parameters (Section~\ref{sec:challenges}). Onboarding CMP202 and PSY555 required only the population of two course-specific artefacts (the RAG library and the KC model); the orchestration codebase was unchanged (Section~\ref{sec:onboarding-pipeline}). 

The RAGAS-based evaluation (Section~\ref{sec:scalability-eval}) supports this at a quantitative level for the components closest to the orchestration layer: Answer Relevancy and Context Precision, both less directly shaped by the generation prompt than Faithfulness, remained stable across all three courses regardless of domain or academic level (0.875--0.935 and 0.877--0.901 respectively).

Two findings complicate a simple yes. First, the course-blind code-question generation logic (Section~\ref{sec:challenges}) encoded assumptions specific to CMP511 and failed outright when applied to PSY555 and CMP202, requiring an explicit fix before the architecture's course-agnostic claim held in practice. Second, the RAGAS Faithfulness gradient (0.685 for CMP511 to 0.499 for PSY555) shows that even components that did not fail outright can degrade as deployment moves further from the system's original design domain. A further complication concerns how domain distance itself should be measured. CMP202's Context Recall (0.404) was markedly lower than either CMP511's (0.609) or PSY555's (0.556), despite CMP202 sharing CMP511's broad disciplinary classification (computing) and PSY555 being, by that classification, the more distant domain from CMP511. This is consistent with the chunking strategy underlying \textit{LEA}'s RAG pipeline degrading specifically for code-dense material (Section~\ref{sec:scalability-eval}) rather than for non-STEM content, suggesting that content structure, not disciplinary class, may be the more useful predictor of where cross-course scalability assumptions break down.

The orchestration architecture itself is course-agnostic by design and this was operationally confirmed; but implementation-wide course-agnosticism is not automatically inherited by every component that touches course-specific data, and surfacing that gap required both deploying a genuinely different course and applying systematic, metric-based evaluation rather than functional testing alone. This section and Section~\ref{sec:scalability} illustrate three distinct ways \textit{LEA} components responded to cross-course deployment: components that are \emph{architecturally invariant} (the orchestration formulae, mastery hierarchy, and RAG pipeline structure), components that are \emph{domain-adapted by design} (the KC model's skill-category taxonomy, built to span procedural and non-procedural objectives), and components where \emph{latent single-course assumptions} were exposed only once a genuinely different domain was deployed against them (the code-question generation logic).

\section{\uppercase{Conclusion}}
\label{sec:conclusion}
This paper extended \textit{LEA}, a tri-modal agentic AI tutoring system, beyond the simulation-only, single-course validation of the conference paper, addressing the two deployment questions it had deferred. We reported the first classroom deployment of \textit{LEA} with real students ($n=8$, CMP511) and the first empirical test of its cross-course scalability claim, deploying the system across three courses spanning two academic levels and two disciplinary domains---including PSY555, a non-STEM social-science course with zero procedural learning objectives---and evaluating retrieval and generation quality across all three using the RAGAS framework.

Together these evaluations show that synthetic simulation was predictively accurate for simple interactions but neither sufficient to anticipate the complexities of real human use nor to detect a production defect that surfaced only in deployment (RQ1), and that \textit{LEA}'s core orchestration required no modification to support a non-STEM, non-procedural course while its auxiliary, content-touching logic did not scale as cleanly (RQ2), with content structure---specifically code-density---proving a better predictor of scalability risk than STEM/non-STEM status.

These findings carry a practical implication: course-agnostic orchestration is achievable and, on this evidence, robust, but any component that touches course-specific content directly (worked-example generation, prompt templates, chunking parameters) should be audited for hidden assumptions before a new domain is onboarded, and simulation should be treated as a filter for refining functionality rather than a substitute for deployment testing.

These findings advance the agenda articulated by Mannekote et al.~\cite{mannekote2024large}: moving beyond prompt-response interaction and synthetic-only evaluation toward systems validated against the complexities of real learners and real deployments. \textit{LEA} has transitioned, over the course of this work, from a simulation-validated proof-of-concept to a classroom-deployed, multi-course platform with human evidence of usability and trust, architectural evidence of cross-course scalability, and quantitative evidence delineating precisely where that scalability claim holds without qualification and where it requires refinement, and including a methodological by-product that re-purposes retrieval metrics as a curriculum-coverage audit as a direction for future work.

\subsection{Limitations}
\label{sec:limitations}

Four limitations bound the conclusions that can be drawn from this work. First, the exploratory CMP511 study ($n=8$, 20\% response rate) precludes inferential analysis; low session engagement means findings reflect initial impressions rather than sustained use, and learning-outcome claims are accordingly conservative (Section~\ref{sec:human_study}). Second, the cross-course RAGAS evaluation used the system's own generated answers as a stand-in reference for Context Precision and Context Recall in the absence of human-authored ground truth; these two metrics should be read as measures of internal retrieval consistency rather than alignment with an independent correctness standard, though Faithfulness and Answer Relevancy are unaffected by this constraint. Third, the hypothesised link between CMP202's sentence-bounded chunking strategy and its lower Context Recall was not isolated through controlled ablation, and other corpus-level differences cannot be ruled out (Section~\ref{sec:scalability-eval}). Finally, the three-course deployment, while spanning two disciplinary domains and two academic levels, remains narrow in absolute breadth.

\subsection{Future Work}
This work motivates six directions.

\begin{itemize}
  \item \textbf{Investigating the chunking-strategy hypothesis.} Re-chunking CMP202's corpus with a code-aware strategy~\cite{zhang2025cast} and re-running the RAGAS evaluation would constitute a controlled ablation isolating chunking strategy as the causal factor in the observed Context Recall deficit, testing whether content structure rather than disciplinary domain drives the degradation identified in Section~\ref{sec:scalability-eval}.

  \item \textbf{RAG corpus--curriculum coverage auditing.} The cross-course evaluation generated its question set from each course's KC model rather than from the RAG corpus, so that every question mapped to a stated learning objective (Section~\ref{sec:scalability-eval}). A by-product of this design is that Faithfulness becomes sensitive to \emph{corpus coverage}: objectives for which the uploaded materials contain no supporting content yield correct-but-ungrounded answers. Turning this confound into an instrument, a systematic comparison of RAGAS performance on KC-derived questions against performance on questions synthesised directly from the indexed corpus (as in standard RAGAS test-set generation~\cite{es2024ragas}) would localise, per learning objective, where a course's corpus fails to cover its stated curriculum. Such a \emph{corpus-adequacy audit} could help instructors identify and fill coverage gaps when authoring or refining a course, before those gaps affect learners. To our knowledge, RAG evaluation metrics have not previously been repurposed in this direction.

  \item \textbf{Auditing remaining course-specific assumptions.} The code-question generation logic and the Faithfulness gradient were two instances of a broader pattern; a systematic audit of fallback content, default values, and example generation throughout the production codebase would identify whether similar latent assumptions persist elsewhere, before a fourth or fifth course is onboarded.

  \item \textbf{Human-authored reference answers.} Collecting instructor-validated reference answers for a sample of questions per course would allow Context Precision and Context Recall to be recomputed against genuine ground truth.

  \item \textbf{Longitudinal and higher-powered classroom evaluation.} A larger, longer-duration pilot, ideally spanning a full term from the start of teaching, would address the sample size and engagement-depth limitations identified in Section~\ref{sec:study-limitations}, and could be extended to CMP202 and PSY555 cohorts to directly test the mode-usage and scaffolding-adaptation questions left open in Section~\ref{sec:scalability-eval}.

  \item \textbf{Pilot-informed refinement of the orchestrator equations.} The conference paper noted that bidirectional retrieval from historical learner interactions and personalised parameterisation of the cognitive-load model's $\beta$ coefficients (currently fixed via offline Monte Carlo simulation, (Eq.\eqref{eq:cognitive_load}) remain open issues. The pilot (Section~\ref{sec:pilot-findings}) highlights two specific targets: the Cognitive Load formulation (Eq.\eqref{eq:cognitive_load}) may need to weight recent struggle more heavily to address the scaffolding rigidity, and the Motivation State formulation (Eq.\eqref{eq:motivation_change}) may require personalised affirmative feedback to mitigate affirmation-fatigue. Both require iterative refinement rather than fixed population-level formulas.

  \end{itemize}

\bibliography{ref}
\end{document}